\documentclass[prb,twocolumn,showkeys,showpacs,preprintnumbers,amssymb,amsmath,floatfix]{revtex4}

\usepackage{graphicx}
\usepackage{bm}
\usepackage{dcolumn}

\begin{document}

\title{Improved modelling of liquid GeSe$_2$: the impact of the exchange-correlation functional}

\author{Matthieu Micoulaut$^{(1)}$}
\author{Rodolphe Vuillemier$^{(1)}$}
\author{Carlo Massobrio$^{(2)}$}

\affiliation{
$^{(1)}$Laboratoire de Physique Th\'eorique de la Mati\`ere Condensée,
Universit\'e Pierre et Marie Curie, Boite 121,
4, Place Jussieu, 75252 Paris Cedex 05, France}
\affiliation{
$^{(2)}$Institut de Physique et de Chimie des Mat\'eriaux de Strasbourg,
23 rue du Loess, BP43,  F-67034 Strasbourg Cedex 2, France}

\date{\today}

\begin{abstract}
The structural properties of liquid GeSe$_2$ are studied by using first-principles
molecular dynamics in conjuncton with the Becke, Lee, Yang and Parr (BLYP)  
generalized gradient approximation for the exchange and correlation
energy. The results on partial pair correlation functions, coordination numbers,
bond angle distributions and partial structure factors are compared with available
experimental data and with previous first-principle molecular dynamics results
obtained within the Perdew and Wang (PW) generalized gradient approximation for the
exchange and correlation energy. We found that the BLYP approach substantially improves upon the
PW one in the case of the short-range properties. In particular, the Ge$-$Ge pair correlation
function takes a more structured profile that includes a marked first peak due to homopolar bonds,
a first maximum exhibiting  a clear shoulder and a deep minimum, all these features being
absent in the previous PW results. Overall, the amount  of tetrahedral order is significantly increased,
in spite of a larger number of Ge$-$Ge homopolar connections.
Due to the smaller number of miscoordinations, 
diffusion coefficients obtained by the present BLYP calculation are smaller by at least
one order of magnitude than in the PW case.
\end{abstract}



\maketitle
\section{Introduction}
Disordered Ge$_{n}$Se$_{1-n}$ materials feature a large variety of bonding
behaviors as a function of the composition.\cite{tronc,azou,liu,boolrev,highT} 
In the liquid state, a delicate interplay between the covalent and the ionic characters
sets in for increasing values of $n$. At $n$=0.33, this results in a network system
(GeSe$_2$) made of predominant GeSe$_4$ tetrahedra coexisting with homopolar bonds and defective
Ge$-$Se coordinations.\cite{petri2} The existence of chemical disorder in an otherwise prevailing
tetrahedral network has been firmly established through the measurement
of the full set of partial structure factors and pair distribution functions of liquid GeSe$_2$
($l$-GeSe$_2$ hereafter).\cite{pensal}
Early molecular dynamics models based on interatomic potentials
were unable to predict miscoordinations and homopolar bonds.\cite{vashcor}
Very recently, a refined model potential for disordered GeSe$_2$
was made available. \cite{mauro1,mauro2} 
This potential reproduces qualitatively the experimental data on glassy GeSe$_2$,
including the presence of Se$-$Se homopolar bonds.
However, Ge$-$Ge homopolar bonds were absent within this description.
This proves that the explicit account of the electronic structrure  in the expression
of the interatomic forces is crucial to describe properly disordered GeSe$_2$ networks. 
Along these lines, two approaches based on density functional theory (DFT)
have been employed to study $l$-GeSe$_2$ via molecular dynamics. D. Drabold and coworkers have
adopted a DFT framework based on a nonself-consistent electronic structure
scheme, the local density approximation of DFT and a minimal basis set.\cite{drabolda,draboldl}
As an alternative, the fully self-consistent evolution of the electronic structure
described within DFT (i.e. first-principles molecular dynamics, FPMD in what follows)
has been pursued in the case of $l$-GeSe$_2$ with the use of plane waves
and pseudopotentials.\cite{carlo1,carlo2,carloprb}
A comparison of the structural properties obtained within these two approaches is provided in a recent paper
for the case of amorphous GeSe$_2$.\cite{carloam}

Turning our attention to the FPMD approach and to the case of $l$-GeSe$_2$, it is worth recalling 
the indications collected through the use of the local density approximation (LDA) within DFT.
This had the effect of producing an atomic structure
affected by an excessive amount of chemical disorder and homopolar bonds.\cite{carlo2}
The absence of the FSDP (first sharp diffraction peak)
in the total neutron structure factor could be correlated to the lack of a predominant
structural unit (the GeSe$_4$ tetrahedron), with comparable percentages of Ge atoms
two-fold, three-fold, four-fold and five-fold coordinated. \cite{carlo2}
Interestingly, interatomic potentials
featuring formal charges on the Ge and the Se atoms (+4 and -2, respectively) are able 
to provide a GeSe$_2$ network based on undefective GeSe$_4$ tetrahedra, together
with a FSDP in the total neutron structure factor. \cite{vashcor} 
In the search of a DFT scheme able to recover a network structure featuring the predominant
presence of GeSe$_4$ tetrahedra, the generalized gradient approximation (GGA) for the exchange and correlation
energy proposed by Perdew and Wang (PW hereafter) was adopted.\cite{pw92} 
This choice yields a very good agreement with experiments 
for the total neutron structure factor over the entire range of momentum transfer.\cite{carlo2}
The improvements brought about by the GGA in the PW form were found to be due to
a better account of the ionic character of bonding, as shown in Ref.\ \onlinecite{carlo2}
through an analysis of the contour plots for the valence charge densities.
The larger ionicity of bonding introduced  by the  PW approach manifests itself through a
larger depletion of the valence charge at the Ge sites and a larger accumulation around
the Se atoms, the covalency remaining essentially equivalent in the LDA
and in the PW scheme.\cite{carlo2}

 Despite this success, detailed comparison of
the partial correlations revealed the existence of residual differences between theory and expriment.
The most important of these differences concerns the first sharp diffraction peak in the
concentration-concentration structure factor that appears in the experiment but it is absent
in our level of theory.\cite{carloprb} In Ref.\ \onlinecite{carloprb}
these shortcomings were attributed to an insufficiently accurate
description of Ge$-$Ge correlations. This was confirmed by the shape of the calculated Ge$-$Ge correlation function,
much less structured than its experimental counterpart and by the excessively long (15 \% more than
the experimental value) first-neighbors Ge$-$Ge distances.
Longer interatomic Ge$-$Ge distances and less structured Ge$-$Ge pair correlation
functions were correlated to an  overestimate of the metallic character in liquid
GeSe$_2$. This observation can be taken as a guideline for the choice of alternative
GGA recipes capable of further improving the performances of DFT and, in particular,
the distribution of the valence charge densities along the bonds. 
In what follows, we are interested in GGA functionals enchancing a $localized$
distribution of the valence electrons at the expenses
of a  $delocalized$ one, intrinsic in schemes inspired by the uniform electron
gas model, as the PW one. 

With this purpose in mind, we present a new set of structural data for short and intermediate range properties
in $l$-GeSe$_2$, based on the GGA scheme after Becke (B) for the exchange energy) and
Lee, Yang and Parr (LYP) for the correlation energy.\cite{becke,lyp}
This recipe makes no assumption on the uniform electron gas character of the
correlation energy. Therefore, it is a good candidate to correct part of the drawbacks
of the PW scheme, as those arising from an overestimate of the metallic character
of bonding.    
The comparative analysis carried out with available PW data reveals that
short-range properties and the diffusion behavior compare significantly better with experiments
within the the BLYP approach.
The improvement is less significant for the intermediate range properties.

This paper is organized as follows. In Sec.~II, we describe our theoretical model.
Our results are collected in two sections, devoted to real
space properties (Sec.~III) and reciprocal space properties (Sec.~IV). 
A summary of general considerations on  the modelling of disordered network-forming
material can be found in Sec.~V. 
Conclusive remarks are collected in in Sec.~VI.

\section{Theoretical Model}

Our simulations were performed at constant volume on a system consisting of 120 atoms
(40 Ge and 80 Se). We used a periodically repeated cubic cell of size 15.7~\AA, corresponding
to the experimental density of the liquid at T$=$1050 K. We refer to our previous results
on $l$-GeSe$_2$ for an extended rationale on the choice of our system size.\cite{carloprb,carlozz2}
The electronic structure was described within density
functional theory (DFT) and evolved self-consistently during the motion.\cite{car85}

In a series of previous papers, we had adopted the PW scheme due to Perdew and Wang
as a generalized gradient approximation.\cite{carlo1,carlo2,carloprb,carloam,carlozz2,carlo2001,carlofsdp}
This amounts to go beyond the local density approximation (LDA)
by using an analytic representation of the correlation energy $\varepsilon$$_c$($\rho$) for a uniform electron gas.
This representation allows for variations of $\varepsilon$$_c$($\rho$) as a function of $\rho$ and
the spin polarization.\cite{pw92} As mentioned in the Introduction, the use of the PW scheme    
improves substantially upon LDA the description of both short and intermediated range order in $l$-GeSe$_2$.
However, residual deficiences are found within the PW when focusing on the Ge$-$Ge correlations, in particular
at the level of the Ge$-$Ge interatomic distance and pair correlation function,
These signatures point out an overestimate of the metallic character of bonding, implying that
the ionicity at the very origin of the formation of the tetrahedral GeSe$_4$ coordination is  undermined 
by an unphysical accumulation of valence electron along the bonds.  

In the search of a GGA recipe correcting these defects, we resort in this work to 
the generalized gradient approximation after Becke (B) for the
exchange energy and Lee, Yang and Parr (LYP) for the correlation energy.\cite{becke,lyp}
It has to be reminded that the record of reliability of the BLYP approach has been firmly
assessed, as shown in a detailed comparative study of the performances of a large
variety of density functional methods. \cite{joblyp} In Ref.\ \onlinecite{joblyp},
it was concluded that the BLYP method is the DFT method with the best overall performances
for properties encompassing equilibrium geometries, vibrational frequencies and atomization
energies of nanostructured systems. 
Furthermore, our choice is motivated by the consideration that no explicit reference to the uniform electron
gas is made in the derivation of the LYP correlation energy. In the original LYP derivation,
a correlation energy formula due to Colle and Salvetti is recast in terms of the electron density
and of a suitable Hartree-Fock density matrix, providing a correlation energy and a  correlation
potential.\cite{colle,lyp} This scheme is expected to enhance the localized behavior of the electron density
at the expenses of the electronic delocalization effects that favor the metallic character.
These effects are built in GGA recipes having the uniform electron gas as reference system,  
as the PW one.
For this rationale to be generally applicable,
it would be desirable to  understand which systems and bonding situations are likely
to be better described by BLYP than by PW. By focusing
on multicomponent systems A$_n$B$_{(1-n)}$ of concentrations $n$,
the BLYP scheme is expected to be  more suited than the PW one to treat bonding
situations characterized by a moderate difference of electronegativity
(termed $\Delta$$_{el}$ hereafter) among the system components. This is especially true
for compositions close to the stoichiometry, to be intended in this context as
the composition at which optimal coordination between the species A and B occurs (for instance
GeSe$_2$ within the Ge$_n$Se$_{(1-n)}$ family).
In the case of a large $\Delta$$_{el}$,
the ionic contribution to bonding is sufficiently large to ensure  effective charge transfer  
in a way essentially independent on the details of the exchange-correlation functional.
This is the case of disordered SiO$_2$ ($\Delta$$_{el}$=1.54),
well described as a corner-sharing network within LDA.\cite{sio2}
On lowering $\Delta$$_{el}$, the amount of the valence charge density along the
bonds becomes non-negligible and the relative weight of the ionic and covalent character
more delicate to quantify. The case of $l$-GeSe$_2$ is a prototype of this situtation,
being characterized by $\Delta$$_{el}$=0.54. By choosing BLYP to improve upon PW (and LDA)
one attempts to mimimize electronic delocalization effects that are undesirable in any
bonding situation characterized by competing ionic and covalent contributions.   
These ideas withstanding, we stress that the use of a GGA recipe localizing
the valence electron density on the atomic sites cannot be legitimated  by the simple consideration   
of the $\Delta$$_{el}$ value. Indeed, it depends also crucially on the variation of bonding
with thermodynamic parameters as pressure and temperature. For instance, in   the
case of $l$-GeSe$_2$ at high temperatures, the delocalized character of bonding grows 
at the expenses of the localized one (due to gap closing effects), thereby making
much less crucial the use of any functional localizing
the electronic valence on the atomic sites.\cite{highT}

In our work, valence electrons were treated explicitly, in conjunction with normconserving
pseudopotentials of the Trouiller-Martins type to account for core-valence interactions.\cite{tm}
The wave functions were expanded at the $\Gamma$ point of the supercell on a
plane wave basis set defined by an energy cutoff of 20 Ry.
We found  bond lengths ($d_0=3.99$ bohr) and vibrational frequencies ($\omega=398$ cm$^{-1}$),
reproducing the experimental data\cite{herz} and the corresponding
PW results quoted in Ref.\ \onlinecite{carloprb}  to within at most 2~\% and 4~\%, respectively.

\begin{figure}
\includegraphics[width=0.9\linewidth]{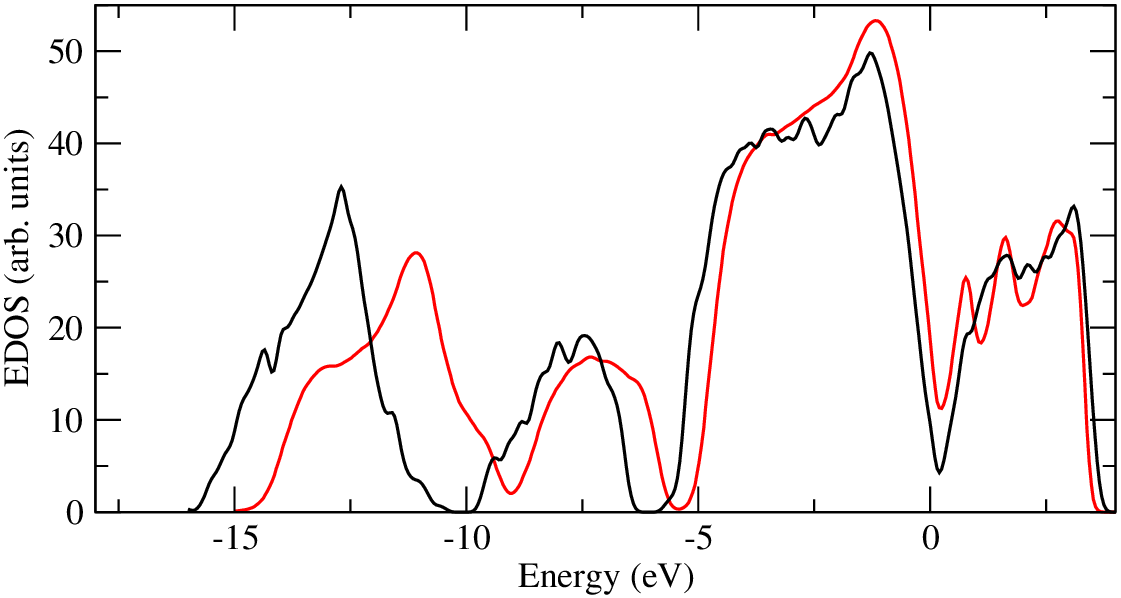}
\vspace{4 cm}
\caption{(color on line) Electronic density of states (Kohn-Sham eigenvalues) of liquid
GeSe$_2$: black line: present BLYP results, red line: PW results
of Ref.\ \protect\onlinecite{carloprb}. A gaussian broadening of 0.1 eV has been employed.}\label{fig0}
\end{figure}

One configuration extracted from the fully equilibrated trajectories obtained
in Refs.\ \onlinecite{carloprb} for liquid GeSe$_2$ with $E_c=20$ 
was taken as initial set of coordinates for the present, new set of calculations
carried out within the BLYP scheme.
We used a fictitious electron mass of 600 a.u.
(i.e. in units of $m_ea_0^2$,
where $m_e$ is the electronic mass and $a_0$ is the Bohr radius)
and a time step of $\delta t=0.1$ fs to integrate the equations of motion.
Temperature control is implemented for both ionic and electronic degrees
of freedom by using Nos\'e-Hoover thermostats.\cite{nose,blochl}
We carried out simulations at T$=$(1050$\pm$10)~K over a time periods of
40 ps, with statistical averages taken after discarding an intial segment of 2 ps.
Since the use  of the BLYP scheme is  aimed at weakening the metallic character of bonding
when compared to the PW one, it is worthwhile to check whether the comparison of the
electronic densities of states for the PW and BLYP case sunstantiate this choice. 
As shown in Fig.\ \ref{fig0}, our selection of the exchange-correlation functional
is legitimated by the comparison between
the corresponding time averages (taken at the same temperature T$=$(1050$\pm$10)~K)
of the electronic densities of states. Indeed, the BLYP approach
is seen to provide a deeper pseudo-gap around the Fermi level than the PW one.

\section{Real space properties}
\subsection{Pair correlation functions}

\begin{figure}
\includegraphics[width=0.8\linewidth]{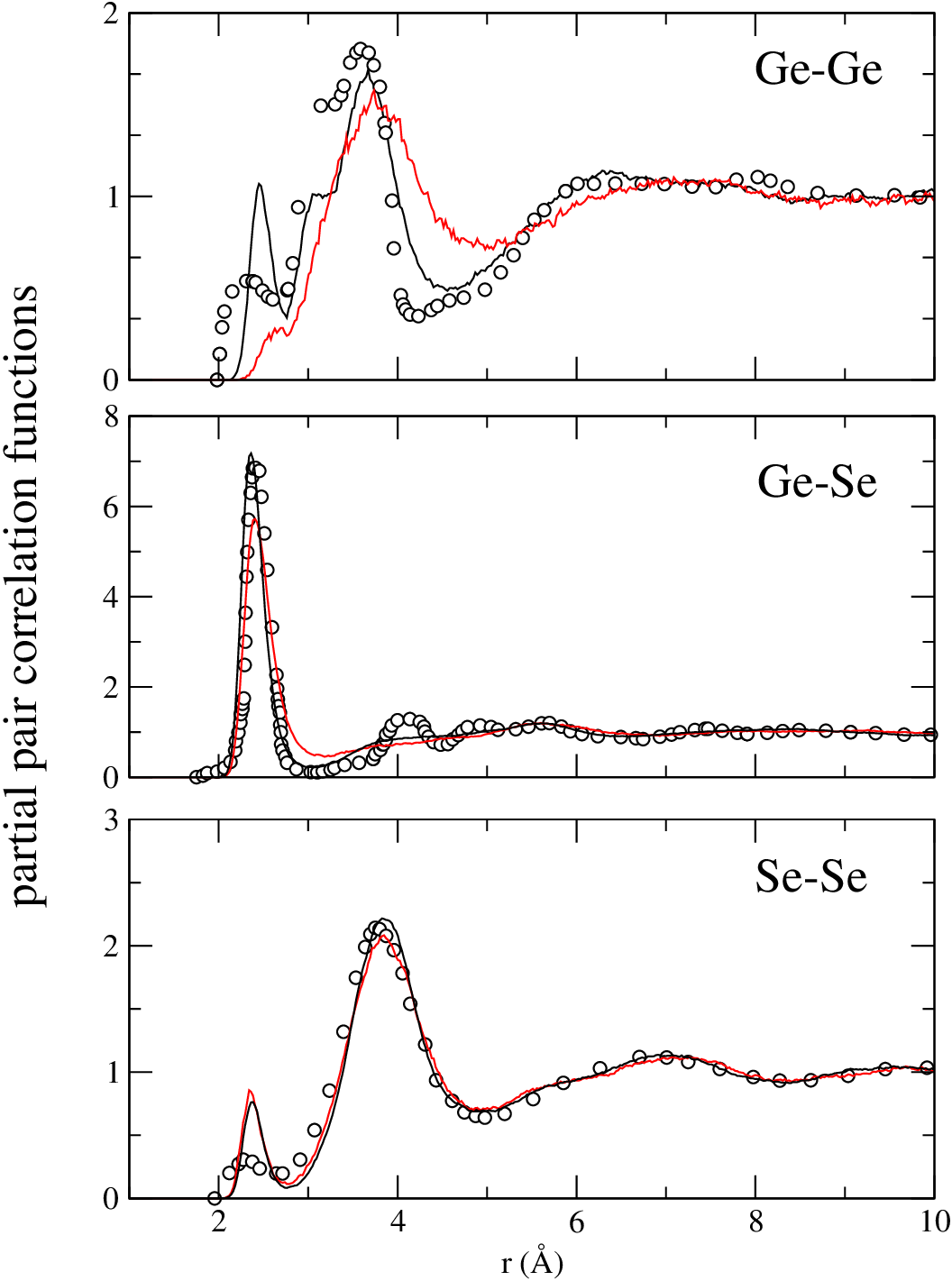}
\caption{(color on line)
Partial pair correlation functions for liquid GeSe$_2$:
black line: present BLYP results, red line: PW results of Ref.\ \protect\onlinecite{carloprb},
open circles: experimental results of Ref.\ \protect\onlinecite{pensal}.
} \label{fig1}
\end{figure}

Partial pair correlation functions $g_{\alpha\beta}(r)$ are shown in Fig.\ \ref{fig1}.
Peak positions and number of neighbors within given integration ranges
are displayed in Table~\ref{table1}.
Among the three pair correlation functions, $g_{GeGe}(r)$ is the one most affected by
the choice of the exchange-correlation functional. The BLYP scheme improves upon the
PW one by yielding a clear first maximum, due to homopolar Ge$-$Ge bonds,
and a very pronounced first minimum, closely reproducing the trends observed in
$g_{GeGe}^{exp}(r)$. In $g_{GeGe}^{BLYP}(r)$ the position of the first peak approaches
the experimental value ($r=2.45$ \AA, BLYP; $r=2.70$ \AA,
PW, Ref.\ \onlinecite{carloprb} ; $r=2.33$ \AA, Ref.\ \onlinecite{petri2}).
Equally favorable is the BLYP prediction of the number of Ge
in the first-neighbor shell (0.22, see Table~\ref{table1},
to be compared with 0.25, Ref.\ \onlinecite{petri2})
The shape of $g_{GeGe}^{BLYP}(r)$ reproduces the shoulder 
in the main peak occurring at $r$$\sim$3.1\AA,
indicative of edge sharing (ES) connections among tetrahedra.\cite{pensal}
The Ge$-$Se pair correlation function $g_{GeSe}^{exp}(r)$ is characterized by a prominent
main peak and a deep minimum. The position of the main peak is slightly displaced
toward shorter distance, by 0.05 \AA. The BLYP scheme is able
to reproduce accurately the height of the first maximum and the abrupt decay from the first,
sharp maximum down to vanishing values.
The first shell of coordination has a number of neighbors (3.55) in
very good agreement with experiments (3.50, see Table~\ref{table1}). Little improvement
is found for larger distances, both $g_{GeSe}^{BLYP}(r)$ and $g_{GeSe}^{PW}(r)$ lacking
of the second, small maximum visible in $g_{GeSe}^{exp}(r)$.
Both  $g_{SeSe}^{BLYP}(r)$ and $g_{SeSe}^{PW}(r)$ follows closely the experimental
$g_{SeSe}^{exp}(r)$ for $r>3$ \AA.
The present set of calculations provides Se$-$Se correlations very
similar to those obtained in Ref.\ \onlinecite{carloprb}, as shown by the close numbers for the
first coordination shell neighbors (see Table~\ref{table1}). Homopolar Se$-$Se bonds
are found at a distance only 2\% larger than in the PW case.

We obtain partial ($n_{Ge}$, $n_{Se}$) and average ($n$)
coordination numbers from the first-neighbor coordination numbers,
$n_{GeGe}$, $n_{GeSe}$, and $n_{SeSe}$, given in Table \ref{table2}.
The resulting theoretical values are compared to experimental data in Table~\ref{table2}.
As pointed out in Ref.\ \onlinecite{carloprb}, a compensation occurred in the PW case
between the underestimated value of $n_{GeGe}$ and the overestimated value of
$n_{GeSe}$, leading to a good agreement for $n_{Ge}$. In the present case,
the small difference between $n_{Ge}$(BLYP) and its experimental counterpart
is the result of close values for each single contributions, i.e. $n_{GeGe}$, $n_{GeSe}$, and $n_{SeSe}$.
As a result, the calculated and experimental average coordination numbers $n$ differ
by only 3.5\%.

Overall, the present  calculations  improve the short-range structure of
$l$-GeSe$_2$, featuring more structured $g_{GeGe}^{exp}(r)$ and $g_{GeSe}^{exp}(r)$.
In particular, the BLYP approach provides much shorter Ge$-$Ge distances and a well
defined first shell of Ge neighbors, bringing pair correlation functions
in better agreement with experiments.

\subsection{Coordination numbers and bond angle distribution}
Further insight into the network topology and its sensitivity to the specific 
exchange-correlation functional can be obtained through $n_{\alpha}{(l)}$.
We define this quantity as the average number of atoms of species
$\alpha$ $l$--fold coordinated (see Table \ref{table3}),
where $\alpha$ are Ge or Se atoms.
Consistently with the choice made in Ref.\ \onlinecite{carloprb}, we here used a cutoff distance of 3~\AA,
which corresponds to the first minimum in the Ge$-$Se pair correlation function
and well describes the first shell of neighbors also for Ge$-$Ge and Se$-$Se correlations.
As a first observation, one notices that the BLYP scheme is characterized by a higher 
proportion of Ge-GeSe$_3$ connections (as much as 23\%), contributing to
a percentage  of Ge fourfold coordinated atoms moderately larger than in the PW case
(66\% against 61\%). The increase of Ge$-$Ge homopolar bonds and Ge-GeSe$_3$ connections
takes place at the expenses of the undefective GeSe$_4$ tetrahedra, lowering from
53.8\% (PW) to 41.8\% (BLYP). These data are a further manifestation of  an increased
number of Ge$-$Ge first-shell neighbors. The distribution of miscoordinated Ge atoms
is different in the two situations: PW and BLYP favor threefold ($l$=3) and
fivefold ($l$=5) connections, respectively. In particular, in Ge-GeSe$_4$ connections
a homopolar Ge$-$Ge bond is found to coexist with an adjacent tetrahedral arrangement.
Coordination of Se atoms reflects the increase of chemical order occuring within the
BLYP scheme. The number of twofold cooordinated Se atoms becomes more than
10\% larger, mostly due to the predominant Se-Ge$_2$ configuration. A corresponding
decrease in the number of miscordinations (in particular Se-SeGe$_2$, lowering from
13\% to 6.4\% is noticeable in Table \ref{table3}.  

\begin{figure}
\includegraphics[width=0.9\linewidth]{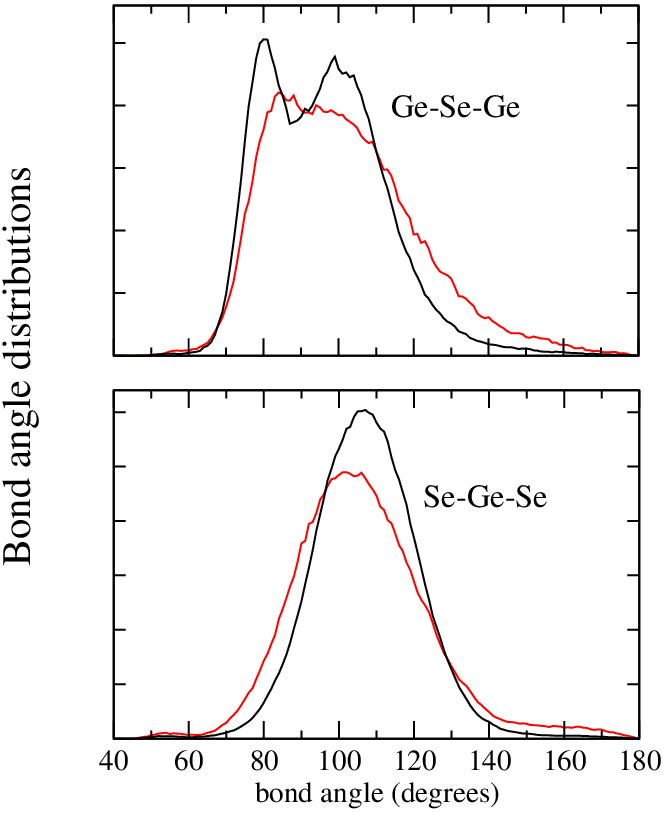}
\vspace{4 cm}
\caption{ (color on line)
Bond-angle distributions Ge$-$Se$-$Ge (top) and
Se$-$Ge$-$Se (bottom). 
black line: present BLYP results, red line: PW results of Ref.\ \protect\onlinecite{carloprb}.
} \label{fig2}
\end{figure}

In~Fig.\ \ref{fig2} we show  the distribution of the Se$-$Ge$-$Se
($\theta_{\rm SeGeSe}$) and Ge$-$Se$-$Ge ($\theta_{\rm GeSeGe}$)
bond angles. These distributions have been calculated by including
neighbors separated by less than 3~\AA.
Two features are worth pointing out. First, the Se$-$Ge$-$Se bond angle distribution 
becomes highly symmetric around 109$^{\circ}$ in the BLYP description and has also
a higher intensity. This stands for an improved tetrahedral order when compared to the
results of Ref.\ \onlinecite{carloprb}, where the maximum occurred at 103$^{\circ}$ and the distribution
has a larger width. Second, two distinct
peaks are visible in the BLYP Ge$-$Se$-$Ge bond angle distribution at about
80$^{\circ}$ and 100$^{\circ}$ at the place of a flat maximum in between the same
values. In Ref.\ \onlinecite{carloprb} we showed that the Ge$-$Se$-$Ge bond angle distribution can be decomposed
in two contributions, one associated to edge-sharing tetrahedra (for values close to 80$^{\circ}$)
and the other due to corner-sharing tetrahedra (for values close to 100$^{\circ}$).
Similarly, the peaks visible in the shape of BLYP Ge$-$Se$-$Ge bond angle distribution have to
be ascribed to these two different connections. Their prominent appearance
stems from a improved tetrehedral organization, in line with the analysis of
the coordination environement for Ge and Se.

Finally, we compare the number of Ge atoms that belong to
zero, one, and two fourfold rings.
We used the counting algorithm based on the shortest-path criterion
first proposed by King and then improved by Franzblau.\cite{king,franzblau,gaito}
Cutoff radii were taken equal to 3.0~\AA\ for Ge$-$Ge, Ge$-$Se and Se$-$Se interactions.
By using the ring statistics results, Ge atoms can be  termed Ge(0) (Ge atoms not
belonging to any fourfold ring),
Ge(1) (Ge atoms belonging to one fourfold ring), and Ge(2) (Ge atoms belonging to
two fourfold rings).
Ge(1) and Ge(2) form edge-sharing connections, while
Ge(0) encompasses not only those Ge atoms involved in
corner-sharing connections ($N_{\rm Ge}$(CS)) but also some of the Ge atoms
forming homopolar bonds, $N_{\rm Ge-Ge}$.
In Ref.\ \onlinecite{carloprb} it was found that 55\% of the Ge atoms do not
belong to fourfold rings (Ge(0)), 36\% belong to a single fourfold ring (Ge(1)), and 9\%
belong to two fourfold rings (Ge(2)). According to  the present results obtained with the BLYP
scheme 61\% of the Ge atoms are Ge(0),  34\% are Ge(1) and  5\% are Ge(2).
Therefore, Ge atoms involved in edge-sharing connections ($N_{\rm Ge}$(ES)),
according to PW and BLYP calculations, turn out to be $N_{\rm Ge}$(ES,PW)= 45\% and
$N_{\rm Ge}$(ES,BLYP) 39\%, respectively. 

To obtain $N_{\rm Ge}$(CS), we adopted the proposal of Ref.\ \onlinecite{petri2},
i.e.\ $N_{\rm Ge}$(CS)= 1$-$$N_{\rm Ge}$(ES)$-$ $N_{\rm Ge-Ge}$,
which holds in the absence of extended chains.\cite{petri2}
By using the results of Table~\ref{table1} ($N_{\rm Ge-Ge}$(BLYP)=22\%, $N_{\rm Ge-Ge}$(PW)=4\%)
one obtaines that $N_{\rm Ge}$(CS, PW)= 51\%
and $N_{\rm Ge}$(CS, BLYP)= 39\%, leading to $N_{\rm Ge}$(CS, PW)/$N_{\rm Ge}$(ES,PW)= 1.13
and $N_{\rm Ge}$(CS, BLYP)/$N_{\rm Ge}$(ES,BLYP)= 1 (these values are also collected
in Table \ref{table3}).
The full partial structure factor analysis of liquid GeSe$_2$
points toward comparable numbers for the  edge-sharing and corner-sharing sites
($N_{\rm Ge}$(CS, BLYP)/$N_{\rm Ge}$(ES,BLYP)$\sim$ 1).\cite{pensal} 
This is consistent with the experimental prediction of Ref.\ \onlinecite{pensal},
where the number of edge-sharing and corner-sharing sites were found very close.
Therefore, the BLYP approach proves more suitable to yield the correct relative
proportion of corner-sharing and edge-sharing tetrahedra.

\section{Reciprocal space properties}
\subsection{Faber-Ziman and Bhatia-Thornton partial structure factors}

\begin{figure}
\includegraphics[width=0.9\linewidth]{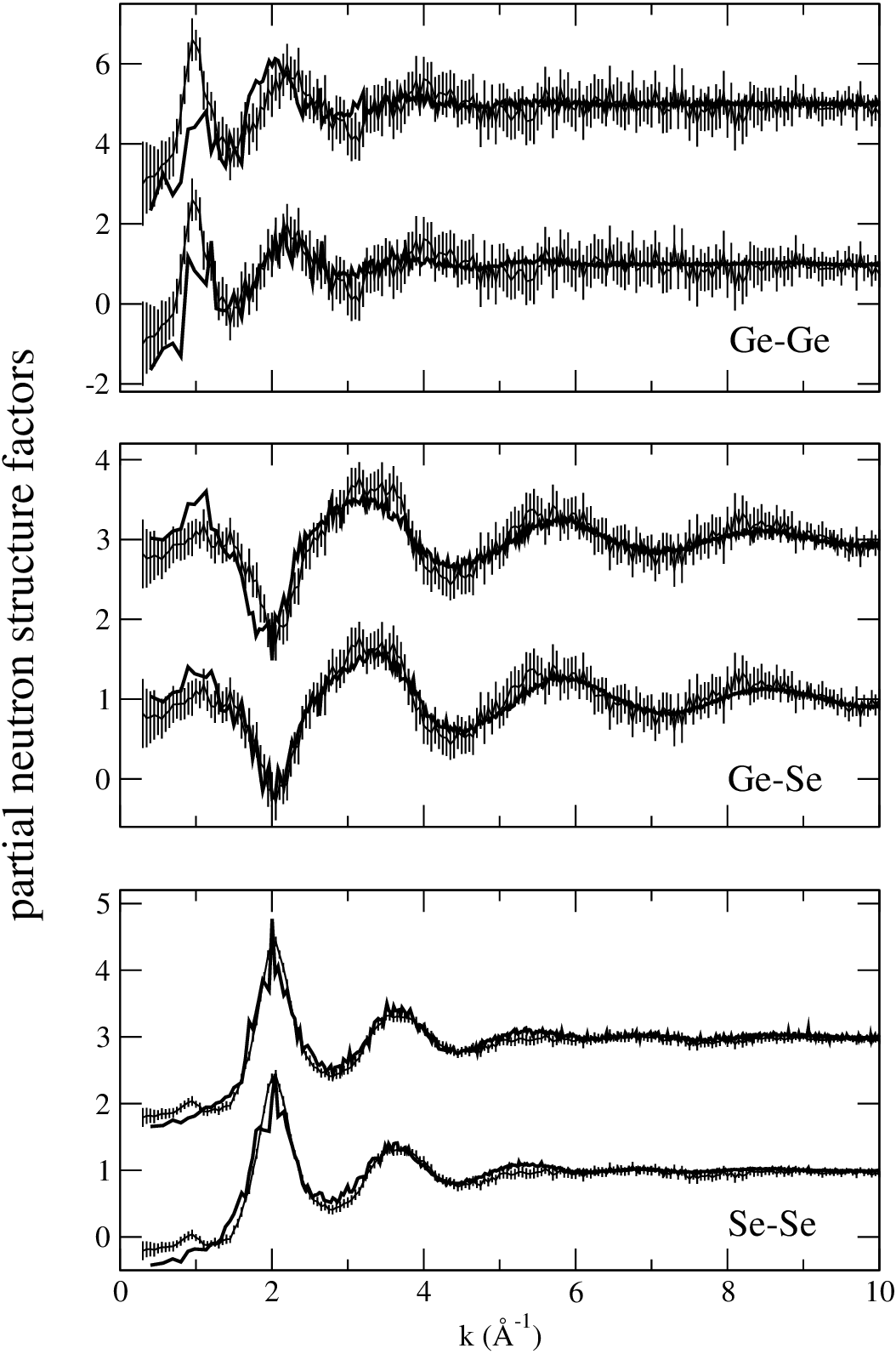}
\caption{
Faber-Ziman partial structure factors for liquid GeSe$_2$.
In each panel, the experimental results of  Ref.\ \protect\onlinecite{pensal} are compared with
the present BLYP results  (bottom part) and with the PW results 
of Ref.\ \protect\onlinecite{carloprb} (top part).
$S_{GeGe}^{PW}(k)$, $S_{GeSe}^{PW}(k)$ and $S_{SeSe}^{PW}(k)$
have been shifted up by 4, 2, and 2, respectively.
 } \label{fig3}
\end{figure}

In Fig.\ \ref{fig3}, we display the comparison between the two sets of calculated 
Faber-Ziman (FZ)\cite{waseda} partial structure factors (PW scheme, Ref.\ \onlinecite{carloprb},
and BLYP scheme, present results) and their experimental counterpart.\cite{pensal,petri2}
In Ref.\ \onlinecite{carloprb}  it was established that the performance of PW calculations  
were very satisfactory for $k$ values characteristic of short range properties ($k>2$ \AA$^{-1}$).
However, a major disagreement existed in the region of the FSDP (first sharp diffraction peak),
at $k$ $\sim$1.~\AA$^{-1}$, in particular for the $S_{GeGe}^{PW}(k)$ structure
factor and, to a lesser extent, for the $S_{GeSe}^{PW}(k)$ structure factor.
Moreover, $S_{GeGe}^{PW}(k)$ was found to be less structured  and 
slightly shifted towards smaller wavevectors with respect to the experimental curve.
For $k$ $>$ 1.5~\AA$^{-1}$ a clear improvement is noticeable in the  
shape of $S_{GeGe}(k)$) around the second maximum, that superposes to the experimental result,
the peak position being located at 2.2~\AA$^{-1}$ (PW calculations, 1.88~\AA$^{-1}$,
experimental value, 2.2~\AA$^{-1}$.\cite{carloprb,pensal}) 
The position of the FSDP in $S_{GeGe}(k)$) is shifted leftward improving upon the
result of Ref.\ \onlinecite{carloprb} in both location and height.
However, a second spurious peak in the FSDP region shows up at larger $k$ values,
suggesting that our statistical accuracy might be further improved by longer runs
and/or larger simulation cells. 
In the case of $S_{GeSe}(k)$, the FSDP height is reduced and the shape of the first
minimum follows closely the experimental profile, bringing theory in
better agreement with the experimental result. 
As to $S_{SeSe}(k)$, both the present approach and the one of Ref.\ \onlinecite{carloprb}
performs very well  over the entire range of $k$ values. 
These pieces of evidence confirm that BLYP concur to improve the short-range behavior
pertaining to Ge atoms, while the changes induces in the intermediate range behavior
are smaller.  

\begin{figure}
\includegraphics[width=0.9\linewidth]{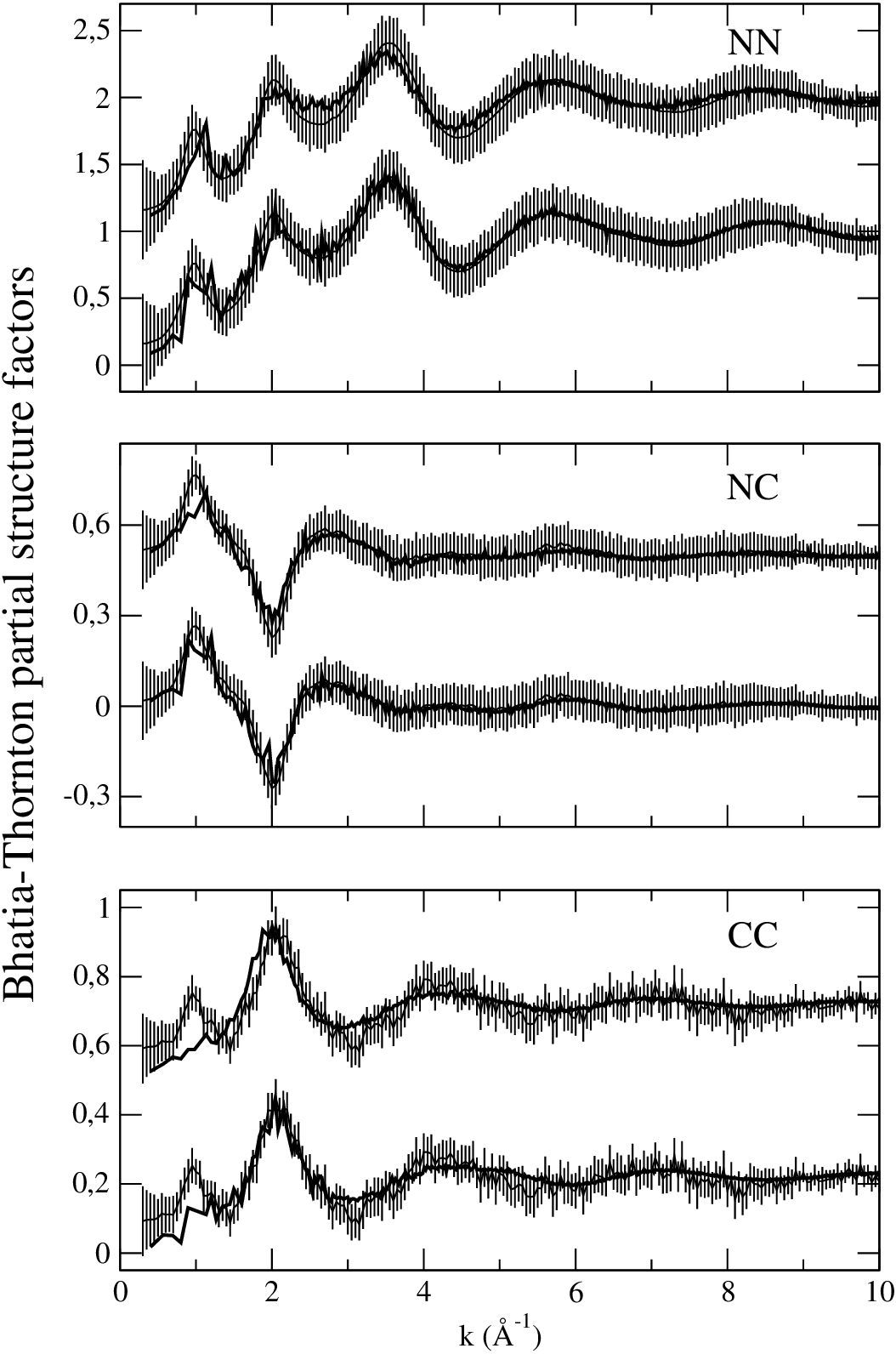}
\caption{
Bhatia-Thornton partial structure factors for liquid GeSe$_2$.
In each panel, the experimental results of  Ref.\ \protect\onlinecite{pensal} are compared with
the present BLYP calculations (bottom part) and with the PW calculations
of Ref.\ \protect\onlinecite{carloprb} (top part).
$S_{NN}^{PW}(k)$, $S_{NC}^{PW}(k)$ and $S_{CC}^{PW}(k)$
have been shifted up by 1, 0.6, and 0.5, respectively.
}
\label{fig4}
\end{figure}

In this context, it is worthwhile to analyze the comparison between
theory and experiments by considering the Bhatia-Thornton\cite{bhatia}
partial structure factors $S_{NN}(k)$ (number-number),
$S_{NC}(k)$ (number-concentration) and
$S_{CC}(k)$ (concentration-concentration) (see Fig.\ \ref{fig4})
These can be obtained by linear combinations of the FZ structure factors.\cite{waseda} 
In terms of the Bhatia-Thornton structure factors, 
the total neutron structure factor $S_{T}(k)$ reads:
\begin {eqnarray}
S_{T}(k)&=& S_{NN}(k) + A\,[{{S_{CC}(k)/c_{{Ge}}\,c_{{Se}}\,}}-1]
 \nonumber\\ &&+B\,S_{NC}(k),
\label{eq1}
\end{eqnarray}
where $A=c_{{Ge}}c_{{Se}}\Delta b^{2}/\langle b \rangle^{2}$,
$B=2\Delta b / \langle b\rangle$, $\Delta b=b_{{Ge}}-b_{{Se}}$,
$\langle b\rangle=c_{{Ge}}b_{{Ge}} + c_{{Se}}b_{{Se}}$,
$c_{\alpha}$ and $b_{\alpha}$ denoting the atomic fraction and
the coherent scattering length of the chemical species $\alpha$
($b_{{Ge}}$=8.185~fm, $b_{{Se}}$=7.97~fm)\cite{pensal}.
This leads to coefficients $A$ and $B$ equal to 1.6$\times$10$^{-4}$ and 0.053, respectively.
As detailed in Ref.\ \onlinecite{carloprb}, $S_{NN}(k)$ is  a very good approximation for the total neutron 
structure factor $S_{T}(k)$, i.e. $\mid$$S_{T}(k)$-$S_{NN}(k)$$\mid$~$<$~0.015.  

As shown in Fig.\ \ref{fig4}, both $S_{NN}^{PW}(k)$ and $S_{NN}^{BLYP}(k)$ are in excellent
agreement with experiments, following closely the experimental data over the
entire range of $k$ values. BLYP improves upon PW in  the 
position of the FSDP, that is closer to the measured value (0.98~AA$^{-1}$, Ref.\ \onlinecite{pensal},
1.01~AA$^{-1}$, present results, 1.13~AA$^{-1}$, Ref.\ \onlinecite{carloprb}) and better reproduces
the experimental data in the range 2~\AA$^{-1}$~$<$~$k$~$<$~4~\AA$^{-1}$. 
Similar performances are also recorded for the case of $S_{NC}(k)$.
Most interesting is the case of the concentration-concentration
structure factor $S_{CC}^{exp}(k)$ in the region $k$~$<$~2.~\AA$^{-1}$.
In recent years, the occurrence  of the FSDP in the concentration-concentration
structure factor $S_{CC}^{exp}(k)$ has stimulated intense experimental and
theoretical work, in the search of its microscopic origins.\cite{salmon92,carlo2001,carlofsdp,sharma}  
With this purpose in mind, the intensities of the FSDP in the $S_{\rm CC}(k)$
have been compared for a series of liquid and glasses.\cite{carlozz2} It was showed that the FSDP
in $S_{\rm CC}(k)$ occurs for moderate departures from chemical order,
but vanishes either when the chemical order is essentially perfect or
for high levels of structural disorder. This is exactly the case of liquid GeSe$_2$
modeled within the PW scheme, for which no FSDP appear in the $S_{\rm CC}(k)$.\cite{carlo1,carloprb}
Given these premises, the question arises on whether the higher tetrahedral
order exhibited by the present model  of liquid GeSe$_2$ 
scheme does result in any improvement in the behavior of the FSDP in $S_{\rm CC}(k)$.
Although a small shoulder has appeared at the FSDP location in $S_{\rm CC}^{BLYP}(k)$,
the prominant experimental peak remains largely underestimated.
The results on the partial structure factor support the notion that  
the use of the BLYP generalized gradient approximation has a much larger effect on
to the short-range properties than on the intermediate-range ones.     
\\

\section{Dynamical properties}

Diffusion coefficients are sensitive probes of the structural organization in 
disordered network-forming materials. On the one hand,  in a chemically ordered network, the tetrahedra
are the main constitutive units, their stability strongly reducing atomic mobility. 
On the other hand, departures from chemical order (homopolar bonds, miscoordinations)
favor atomic mobility since the network can seek the energetically most favorable arrangements
through bond coordination changes. 
In a recent paper, we have provided a revealing example of the correlation existing  between
the values of the diffusion coefficients and the network structure.\cite{mwcm}
In Ref.\ \onlinecite{mwcm}
we have compared structural and dynamical properties of liquid GeSe$_2$ as obtained
from FPMD  and an effective potentials based the polarizable
ionic model (PIM). While a sizeable departure from chemical order is found in the FPMD model,
the PIM structure is highly chemically ordered and the GeSe$_4$ tetrahedron is  largely predominanant among
the structural units. As a consequence of these drastic structural differences, the diffusion
coefficients pertaining to the PIM model are  as low as $\sim{1}\times{10}^{-6}$cm$^2$s$^{-1}$
at T=3000~K, while they are at least ten times larger in the FPMD case at T=1050~K (see below).    
Having established that the tetrahedral order is larger in the BLYP
model than in the PW one for $l$-GeSe$_2$ at T=1050~K, it is of interest to see to what extent
this is reflected in the values of the corresponding diffusion coefficients.

\begin{figure}
\includegraphics[width=0.9\linewidth]{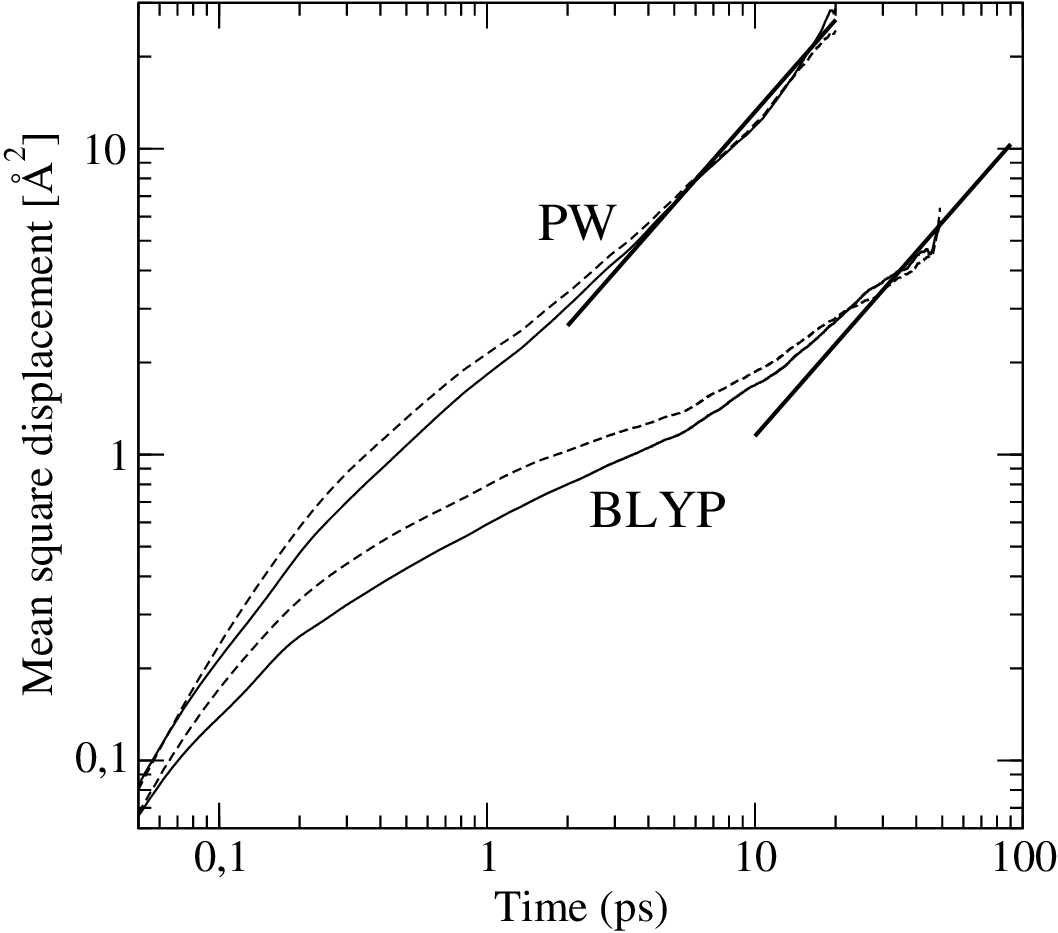}
\vspace{4 cm}
\caption{
Average mean square displacements for Ge (dashed line) and Se atoms
(full line) of liquid GeSe$_2$. The infinite time behavior corresponding
to a slope equal to 1 in the $log$$-$$log$ plot of $<$$r^2$(t)$>$~$vs$~$t$
is also shown.
}
\label{fig5}
\end{figure}

The comparison between the calculated statistical average of the mean square displacement
\begin {equation}
<r^2(t)> = \frac {1}{N_{\alpha}}
{\langle\sum_{i=1}^{{N}_{\alpha}}{|{\bf r}_{i{\alpha}}(t)-{\bf r}_{i{\alpha}}(0)|}^2\rangle}
\label{eq4}
\end{equation}
for both species $\alpha$, Ge and Se, obtained within the PW (Ref.) and the BLYP
schemes, is shown in Fig.\ \ref{fig5}.
In Eq.\ (\ref{eq4}), ${\bf r}_{i\alpha}(t)$ is the coordinate of the $i$th particle at
time $t$ and ${N}_{\alpha}$ is the number of particles of the species $\alpha$.
The diffusive regime is associated with a  slope equal to 1
in the infinite time limit linear behavior of $log$ $<$$r^2$(t)$>$ $vs$ $log$~$t$.
Provided this condition holds, the diffusion constant can be obtained as
\begin{eqnarray}
D={\frac {<r^2(t)>}{6t}}
\end{eqnarray}
In Ref.\ \onlinecite{carloprb}, asymptotic values of $<$$r^2$(t)$>$  are  attained after 5$-$6~ps to give
diffusion coefficients of $D_{{Ge}}$$=$(2.2$\pm$0.2)$\times$10$^{\rm {-5}}$cm$^{\rm 2}$/s
and $D_{{Se}}$$=$(2.2$\pm$0.2)$\times$10$^{\rm {-5}}$cm$^{\rm 2}$/s.
In the present case, establishment of a diffusive regime requires longer
time intervals and asymptotic values of $<$$r^2$(t)$>$ could only be extrapolated
in Fig.\ \ref{fig5}. The estimated diffusion coefficient takes the value of 
$D_{{\alpha}}$=(0.2$\pm$0.2)$\times$10$^{\rm {-5}}$cm$^{\rm 2}$/s.
for both $\alpha$=Ge and Se.
To allow a comparison with experimental data, we use the Eyring relationship;
\begin{eqnarray}
\label{Eyring}
D={\frac {kT}{\eta\lambda}}
\end{eqnarray}
where $\lambda$ is a typical hopping length for the diffusing atom\cite{Eyring} and $\eta$
is the viscosity.
In silicate melts such as $Na_2SiO_3$, Eq. (\ref{Eyring}) holds satisfactorily 
with $\lambda$=2.8~\AA, a distance typical of Si-Si and O-O separation in these melts.\cite{sivisco}
By exploiting viscosity data of liquid $GeSe_2$,
together with a choice of $\lambda$ close to the Ge$-$Ge and Se$-$Se separation distances (3.7~\AA),
we obtain for the temperatures $T=1050~K$ a diffusion coefficient
$D_{{\alpha,exp}}$=0.045$\times$10$^{-5}$~cm$^{\rm 2}s^{-1}$/s.\cite{diffexp}
This estimate can be taken as a lower bound in Eq. (\ref{Eyring}), since the account
of homopolar bonds for both species increases $\lambda$. The present calculations are in better
agreement with the estimated value of $D_{{\alpha}}$ than the results of Ref.\ \onlinecite{carloprb}.
Such improved agreement is consistent with the partially restored tetrahedral order
that characterizes our real space data. Restored tetrahedral order enhances the strenght of the network,
thereby reducing the atomic mobility.

\section{Discussion}
In view of the above pieces of evidence, it appears that the structure of a prototypical network-forming
material, such as $l$-GeSe$_2$, is highly sensitive to the specific exchange-correlation functionals
employed with DFT. In particular, the extent of the departure from perfect chemical order (i.e. no
homopolar bonds and all Ge atoms four-fold coordinated) depends on the spatial distribution
of the valence charge density. In the case of moderate difference of electronegativity between the
component systems, the account of the  delicate balance between electron localization on the atomic sites
(i.e. ionic bonding) and electronic delocalization (i.e. covalent effects and/or tendency toward
metallic bonding) becomes a challenging issue for atomic-scale modelling.     
Liquid GeSe$_2$ proved to be an interesting benchmark systems for theoretical approaches, 
due to the concomitant presence of close percentages of corner- and edge-sharing connections,
homopolar bonds and fluctuations of concentrations on intermediate range distances.
Early effective potentials based on the coulombic interations between formal charges
provide the extreme case of ionic interactions, leading to networks made of
undefective tetrahedra and the absence of homopolar bonds.\cite{vashcor} The inclusion of polarization
effects and/or many body forces allows for the presence of edge-sharing tetrahedra,
in agreement with experimental results.\cite{vashcor,mauro1,mauro2,mwcm} The use of DFT models 
proved necessary to obtain defective tetrahedra (Ge atoms not fourfold coordinated) and homopolar
bonds for both Se and Ge atoms, as found in the experiments. 
However, the mere application of a fully self-consistent LDA had the effect of
overestimating the chemical disorder. This has led to the lack of intermediate range order
and to an overall atomic structure in worse agreement with the experiment than that obtained
with interatomic potentials.\cite{carlo2} These shortcomings were traced back to an insufficient treatment
of the ionicity, i.e. an insufficient valence charge localization on the atomic sites,
strongly affecting the stability of GeSe$_4$ tetrahedra. To go beyond these limitations
of the LDA-DFT approach, we have applied the generalized gradient approximation of DFT
in two subsequent steps. In the first (Ref.\ \onlinecite{carlo2}) a functional rooted on
the electron gas model has allowed to recover a very good agreement with the experiments
in terms of structural properties. In the second (this work) a further refinement aimed
at a more enhanced valence charge localization on the atomic site has improved the
short range properties, in particular those related to the Ge close environment. 

It is useful in this context to recall the role played by specific structural
features in determining a realistic network structure. The following ideas apply to
the whole family of A$_{n}$X$_{1-n}$ (A=Ge, Si; X=O, S, Se) disordered systems.
The structural features we refer to are : {\it {a)}} homopolar
bonds and miscoordinations and {\it {b)}} edge-sharing tetrahedra $vs$ corner-sharing
tetrahedra. Being able to correctly describe the departure from chemical order 
(occurring because of miscoordinations and homopolar bonds)
is crucial for two reasons. First, it allows
to model the short range structure reliably improving upon simplified models
consisting of  perfect tetrahedra connected to each other. Second, it has a profound impact on
the intermediate range properties, which were found to be strongly dependent on the
existence of a moderate amount of chemical disorder. As detailed in a previous paper,
no intermediate range order exist either in perfect tetrahedral or in
highly disordered networks.\cite{carlozz2} 
The establishement of the intermediate range order is also very much sensitive to the
occurrence of edge-sharing tetrahedra. 
Recently, it has been shown that the first sharp diffraction peak in the concentration-concentration
structure factor $S_{\rm CC}(k)$ is due to chains of edge-sharing tetrahedra. \cite{carlofsdp}
This result correlates a specific structural subunit to the intermediate range order,
exemplyifing how an accurate description allows to link 
microscopic features to measurable properties. 

\section{Conclusion}

We have studied the short and intermediate range properties of liquid GeSe$_2$ at T=1050 K
in the framework of first-principle molecular dynamics by using for the  generalized gradient
approximation the BLYP scheme. Our motivation rests on the outcome of a previous investigation,
carried out within the same theoretical framework but
employing a different recipe (the one due to Perdew and Wang, PW) for the generalized gradient
approximation.\cite{carloprb} In that work, it was stressed that the comparison between theory and experiment
was excellent at the level of the total neutron structure factor, but less satisfactory
in terms of the partial correlations both in real and reciprocal space.
We recall that in Ref.\ \onlinecite{carloprb},  
the Ge$-$Ge pair correlation function was less structured than the experimental counterpart,
the first neighbors distances exceeded the experimental values by about 15 \% and no features were found
at the FSDP location in the  concentration-concentration structure factor. 
The use of the BLYP exchange-correlation functional substantially improves the
short-range properties of liquid GeSe$_2$. The shapes of the Ge$-$Ge and (to a lesser extent)
the Ge$-$Se pair correlation functional are more structured and are consistent with
a higher level of tetrahedral organization, i.e. the network is
less affected by coordinations other than the fourfold one (GeSe$_4$ tetrahedra).  
The same occurs for the  Se$-$Ge$-$Se and Ge$-$Se$-$Ge bond angle distributions,
the first more symmetric around the tetrahedral angle 109$^{\circ}$
while the second shows two distinct peaks accounting for edge-sharing and corner-sharing
connections.
Higher tetrahedral order results in lower diffusion coefficient for both species.
The impact of the BLYP scheme on the intermediate range properties is more elusive.   
Smaller improvements are found for the intensity and the position of the peaks
located at low $k$ values in the partial structure factor.
Work is in progress to investigate the impact of hybrid exchange-correlation functionals
on the network properties of disordered chalcogenides.\cite{hybrid}

\section{Acknowledgement}
We thank P. S. Salmon and M. Boero for stimulating exchanges.

\newpage
\begin{table}
\caption{First (FPP) and second (SPP) peak positions in
experimental (Ref.\ \protect\onlinecite{pensal,petri2}) and theoretical $g_{\alpha\beta}$(r).
BLYP: present calculations, PW: calculation of Ref.\ \protect\onlinecite{carloprb}.
The integration ranges corresponding to
the coordination numbers $n_{\alpha\beta}$ and $n'_{\alpha\beta}$ are
0$-$2.6~\AA, 2.6$-$4.2~\AA~for $g_{GeGe}$(r), 0$-$3.1~\AA,
3.1$-$4.5~\AA~for $g_{GeSe}$(r) and 0$-$2.7~\AA, 2.7$-$4.8~\AA~for
$g_{SeSe}$(r).    
Error bars are the standard deviations from the mean for subaverages
of 2~ps.
}
\vspace{1cm}
\begin{ruledtabular}
\begin{tabular}{lllll}
  $g_{\alpha\beta}$(r)&FPP~(\AA)&$n_{\alpha\beta}$&
SPP~(\AA)&$n'_{\alpha\beta}$\\
\\
$g_{GeGe}^{BLYP}$(r)
   &{{2.45$\pm$0.10}}&{{0.22$\pm$0.01}}&{{3.67$\pm$0.10}}&
{{2.70$\pm$0.06}}\\
$g_{GeGe}^{PW}$(r)
   &{2.70$\pm$0.10}&{0.04$\pm$0.01}&{{3.74$\pm$0.05}}&
{2.74$\pm$0.06}\\
$g_{GeGe}^{exp}$(r) 
   &{2.33$\pm$0.03}&{0.25$\pm$0.10}&{3.59$\pm$0.02}&
{2.9$\pm$0.3}\\
&\\
$g_{GeSe}^{BLYP}$(r)
   &{{2.36$\pm$0.10}}&{{3.55$\pm$0.01}}&{{5.67$\pm$0.02}}&
{{3.85$\pm$0.06}}\\
$g_{GeSe}^{PW}$(r)
   &{{2.41$\pm$0.10}}&{{3.76$\pm$0.01}}&{{5.60$\pm$0.01}}&
{{3.72$\pm$0.03}}\\
$g_{GeSe}^{exp}$(r) 
   &{{2.42$\pm$0.02}}&{{3.5$\pm$0.2}}&{{4.15$\pm$0.10}}&
{{4.0$\pm$0.3}}\\
&\\
$g_{SeSe}^{BLYP}$(r)
   &{{2.38$\pm$0.02}}&{{0.33$\pm$0.01}}&{{3.83$\pm$0.02}}&
{{8.9$\pm$0.06}}\\
$g_{SeSe}^{PW}$(r)
   &{{2.34$\pm$0.02}}&{{0.37$\pm$0.01}}&{{3.84$\pm$0.02}}&
{{9.28$\pm$0.04}}\\
$g_{SeSe}^{exp}$(r)
   &{{2.30$\pm$0.02}}&{{0.23$\pm$0.05}}&
{{3.75$\pm$0.02}}&{{9.6$\pm$0.3}}\\
&\\
\end{tabular}
\end{ruledtabular}
\label{table1}
\end{table}

\newpage
\begin{table}
\caption{Experimental and theoretical values for the
partial coordination numbers $n_{Ge}$  and $n_{Se}$ and
the average coordination number $n$ of liquid GeSe$_2$ at T$=$1040~K.
BLYP: present calculations, PW: calculation
of Ref.\ \protect\onlinecite{carloprb}.
The coordination numbers $n_{Ge}$
and $n_{Se}$ are given by $n_{GeGe}$ + $n_{GeSe}$ and
$n_{SeSe}$ + $n_{SeGe}$, respectively (see the values reported in
Table~\ref{table1} for $n_{GeGe}$, $n_{GeSe}$ and $n_{SeSe}$,
where $n_{GeSe}$ = 2$n_{SeGe}$).
The average coordination number $n$
is equal to  $c_{Ge}$($n_{GeGe}$ + $n_{GeSe}$) + $c_{Se}$ ($n_{SeSe}$ + $n_{SeGe}$).
The experimental values extracted from Ref.\ \protect\onlinecite{petri2}
are also reported.
Error bars are the standard deviations of the mean for subaverages
of 2~ps.
}
\vspace{1cm}
\begin{ruledtabular}
\begin{tabular}{llll}
&$n_{Ge}$&$n_{Se}$&$n$\\
BLYP&3.77$\pm$0.02&2.11$\pm$0.02&2.66$\pm$0.02\\
PW&3.80$\pm$0.02&2.25$\pm$0.02&2.77$\pm$0.02\\
Ref.\ \protect\onlinecite{petri2}&3.75$\pm$0.3&1.98$\pm$0.15&2.57$\pm$0.20\\
&\\
\end{tabular}
\end{ruledtabular}
\label{table2}
\end{table}

\begin{table}
\caption{
Average number $n_{\alpha}(l)$ (expressed as a percentage)
of Ge and Se atoms $l$--fold coordinated  at a distance of 3.0 ~\AA.
For each value of $n_{\alpha}(l)$, we give the identity and the
number of the Ge and Se neighbors.
For instance, GeSe$_3$ with $l$=4 means a fourfold coordinated Ge
with one Ge  and three Se nearest-neighbors.
Values smaller than 1 are reported only for sake of
comparison with corresponding values equal or larger than 1.
In parenthesis: results
of Ref.\ \protect\onlinecite{carloprb}.
We also compare calculated and
experimental values (in percentage) for the number of Ge atoms
forming edge-sharing connections, $N_{\rm Ge}$(ES), the number of Ge
atoms forming corner-sharing connections, $N_{\rm Ge}$(CS) and the
number of Ge atoms involved in homopolar bonds, $N_{\rm Ge-Ge}$. 
Experimental values are taken from Ref.\ \protect\onlinecite{petri2}.
}

\vspace{0.3cm}
\begin{ruledtabular}
\begin{tabular}{llllll}
   {\bf Ge}&&$l=2$&&$l=3$&\\
   &&Se$_2$&4.0~(5.2)&GeSe$_2$&0.8~(2.6)\\
   &&&&Se$_3$&13.5~(19.8)\\
   \\
   &&$l=4$&&$l=5$&\\
   &&GeSe$_3$&23.3~(7.0)&Ge$_2$Se$_3$&2.4~(0.4)\\
   &&Se$_4$&41.8~(53.8)&GeSe$_4$&11.7~(5.9)\\
   &&&&Se$_5$&0.6~(4.6)\\
\colrule
\\
   {\bf Se}&&$l=1$&&$l=2$&\\
   &&Ge&0.9~(1.7)&Se$_2$&3.6~(2.8)\\
   &&&&SeGe&20.7~(21.9)\\
   &&&&Ge$_2$&59.2~(45.6)\\
   &&&&&\\
   \\
   &&$l=3$&&$l=4$&\\
   &&Se$_2$Ge&2.4~(3.2)&SeGe$_3$&0.3~(1.0)\\
   &&SeGe$_2$&5.8~(8.6)&&\\
   &&Ge$_3$&6.4~(13.1)&&\\
\\
\colrule
  &&$N_{\rm Ge}$(ES)&$N_{\rm Ge}$(CS)&$N_{\rm Ge-Ge}$&\\
  This work &&39&39&22&\\
  Ref.\ \protect\onlinecite{carloprb} &&45&51&4&\\
\end{tabular}
\end{ruledtabular}
\label{table3}
\end{table}

\vspace{4 cm}

\vspace{6 cm}

\vspace{4 cm}

\vspace{4 cm}

\vspace{4 cm}

\vspace{4 cm}
\end {document}